\documentclass[10pt, conference, compsocconf]{IEEEtran}
\usepackage[, top=0.75in, left=0.62in, right=0.62in, bottom=0.8in, includefoot]{geometry}
\usepackage{amsmath, amssymb, amsthm,algorithm}
\usepackage[latin1]{inputenc}
\usepackage{xcolor}
\usepackage{latexsym}
\usepackage{graphics,epsfig}
\usepackage{amsfonts}
\usepackage{booktabs}
\usepackage{color}
\usepackage{multirow}
\usepackage[justification=centering]{caption}
\usepackage{graphicx}
\usepackage{adjustbox}
\usepackage{kantlipsum}
\usepackage[caption=false,font=footnotesize]{subfig}
\usepackage{cases}
\usepackage{url}
\usepackage{tabu}
\usepackage{makecell}
\usepackage{soul}

\usepackage{array}
\usepackage{algpseudocode}
\usepackage{mathtools,lipsum,cuted}

\usepackage[justification=centering]{caption}

\usepackage[caption=false,font=footnotesize]{subfig}
\usepackage[T1]{fontenc}
\usepackage{mwe}
\usepackage{subfig}

\newlength{\tempheight}
\newlength{\tempwidth}

\newcommand{\rowname}[1]
{\rotatebox{90}{\makebox[\tempheight][c]{#1}}}

\newcommand{\columnname}[1]
{\makebox[\tempwidth][c]{#1}}

\makeatletter
\def\BState{\State\hskip-\ALG@thistlm}
\makeatother

\newcounter{example}[section]

\usepackage[english]{babel}
\usepackage{verbatim}

\usepackage[english]{babel}

\theoremstyle{definition}

\theoremstyle{remark}

\begin{document}
\title{Enhancing REST HTTP with Random Linear Network Coding in Dynamic Edge Computing Environments}
\author{\IEEEauthorblockN{Cao Vien Phung, Jasenka Dizdarevic, Francisco Carpio and Admela Jukan}
\IEEEauthorblockA{Technische Universit\"at Braunschweig, Germany\\
Email: \{c.phung, j.dizdarevic, f.carpio, a.jukan\}@tu-bs.de
}}
\maketitle

\begin{abstract}
The rising number of IoT devices is accelerating the research on new solutions that will be able to efficiently deal with unreliable connectivity in highly dynamic computing applications. To improve the overall performance in IoT applications, there are multiple communication solutions available, either proprietary or open source, all of which satisfy different communication requirements. Most commonly, for this kind of communication, developers choose REST HTTP protocol as a result of its ease of use and compatibility with the existing computing infrastructure. In applications where mobility and unreliable connectivity play a significant role, ensuring a reliable exchange of data with the stateless REST HTTP protocol completely depends on the developer itself. This often means resending multiple request messages when the connection fails, constantly trying to access the service until the connection reestablishes. In order to alleviate this problem, in this paper, we combine REST HTTP with random linear network coding (RLNC) to reduce the number of additional retransmissions. We show how using RLNC with REST HTTP requests can decrease the reconnection time by reducing the additional packet retransmissions in unreliable highly dynamic scenarios.
\end{abstract}

\section{Introduction}\label{introduction}
Machine to machine communication with its ongoing development is considered a key aspect to be studied in the area of the Internet of Things (IoT). IoT scenarios come with a high number of implementation difficulties demanding computation tasks to be performed in different networks and system architectures, all while maintaining high mobility and dynamicity, and dealing with different challenges ranging from resource management, communication and interoperability issues to data processing and analysis. In order to satisfy the requirements of these new scenarios, well known and accepted technologies such as cloud computing, have been merging with novel technologies that are shifting part of the computation closer to the edge devices, known as fog computing. There have been many research efforts and projects dedicated to solving each of the problems found in these scenarios with fog-to-cloud system solutions, many of which are focused on optimizing network infrastructure and connectivity itself. In this paper we will focus on the improvement of the communication aspect of these systems, particularly on the application layer communication in highly dynamic mobile scenarios by combining the REST HTTP protocol with random linear network coding.

The HTTP protocol following the architectural style defined by REST is being widely used as a communication protocol for web services and also for creating REST APIs for distributed system communication. The ease of use and compatibility with the already existing systems made its adoption as a communication protocol faster than with any other protocol, even with the known limitations this protocol has in some scenarios. One of these scenarios is when building RESTful applications with certain reliability requirements in dynamic environments where connectivity is intermittent and unreliable. A common developer practice for dealing with that kind of situations where timeout events occur is to resend request message following certain self made procedure, instead of any standard procedure. Due to the intrinsic nature of REST HTTP as a polling protocol, the so-called unsafe methods can modify resources in the server side even when the acknowledgments fail. This problem makes the client unaware of the modified resources and forces the client to resend repeated requests. In order to avoid duplicated modification of resources, some policies are usually applied on the server side to make the client aware whether the resources were already modified or not.

In this paper we try to address this issue by combining REST HTTP with random linear network coding (RLNC), a coding technique invented by the authors in \cite{1228459}, in order to minimize the amount of extra requests that have to be sent to the server. We propose a solution in form of a library for the developer to use that will automatically perform RLNC over REST HTTP with no extra development effort for the developer. In our coding scheme, instead of sending native messages, we dispatch coded messages, where the main goal is to predict the loss rate and adjust, more accurately, the number of additional messages in order to improve bandwidth utilization. The designed scheme is designed to be applied in dynamic environments, where the communication between client and server is intermittent. Specifically, we study the case where a mobile client, for instance a smart car, wants to update information to different servers located in base stations along a roadway, and because of tunnels the signal is intermittently lost. Our numerical results show how we reduce the number of additional messages necessary for the client to update the data when using network coding in combination with REST HTTP.

\section{Related Work} \label{RelatedWork}

Handling dynamic mobile scenarios has been one of the key issues for many real-time IoT based systems. In \cite{Sharma2017a} authors explain the limitations of cloud computing solutions in handling mobility issues in these kind of systems. As a solution they propuse a framework that combines cloud comuting with computing closer to end devices in a wireless IoT systems.
The advantages of the fog computing in different dynamic IoT application scenarios have been also detailed in \cite{Chiang2016} and \cite{Vilalta2018}. While \cite{Chiang2016} offers a more general overview of these advantages, \cite{Vilalta2018} focuses on a specific scenario which includes communication between smart vehicles and their fog computing nodes positioned at base stations.

However, even with the improvements gained with fog based system architectures, the issue of intermittent connections in highly dynamic IoT applications and disruptions that come as their consequence has still many open questions. This has led to many different research efforts in improving these solutions. In \cite{Gia2018} authors approach the problem by developing a handover mechanism for mobility support in a IoT-fog systems tested in a health monitoring application. The handover procedure has also been optimized for another fog based framework that tackles high dynamic scenario of connected vehicles in \cite{Li2019}. Beside handover optimization, the choice of the application layer protocol has also been a subject of research when tackling consequences of unreliable connections in these kind of solutions. In \cite{Vilalta2018} authors are using a fog based solution and RESTCONF, an HTTP based protocol for smart vehicle related communication and data computations. In \cite{LeSommer2016} authors have presented a disruption-tolerant RESTful support, tested both with HTTP and CoAP. Their main goal was to improve communication in a dynamic scenario where many devices are prone to disconnections while moving. Idea of improving communication with the adaptation of REST can be used, this time by using network coding.

Network coding (NC) can be dated in 2000 \cite{Ahlswede2000}, a technique which allows network systems to combine several native messages into one coded message in order to expand the maximum bandwidth utilization. In \cite{Hundeboll2014} authors use a network coded protocol operating between the network and transport layers in a wireless network. The results have shown that by using RLNC, this protocol was able to recover from packet losses. In order to improve performances of dynamic IoT scenarios the interesting path is the combination of network coding and fog based computing. Possible applications of NC in IoT and fog based systems have been described in \cite{Peralta2018} with promosing results reported in \cite{Marques2017}, where authors have used NC to improve efficiency of data communication protocols in fog computing wireless sensor environment. In this paper we will explore combination of NC and REST HTTP protocol in IoT to fog communication scenario, as it is still application layer protocol of choice for developers according to multiple research efforts as the one reported in \cite{Dizdarevic2018}.

\section{System design}\label{RESTWNC}
This section shows our solution on applying network coding operations as an embedded mechanism on top of the HTTP protocol when using it with REST. Before entering into details, we recall one definition and one proposition for the concept of "Seeing a packet" taken from \cite{Jay2011}:

\textbf{Definition 1} (Seeing a packet): A node is said to have seen a packet $p_k$ if it has enough information to compute a linear combination of the form ($p_k+q$), where $q=\sum_{l>k}\alpha_l p_l$, with $\alpha_l\in F_q$ for all $l>k$. Thus, $q$ is a linear combination involving packets with indices larger than $k$.

\textbf{Proposition 1}: If a node has seen packet $p_k$, then it knows exactly one linear combination of the form $p_k+q$ such that q is itself a linear combination involving only unseen packets.

Based on this assumptions, upon receiving a coded packet, instead of waiting to have enough information to decode the desired packets, the server immediately tries to perform Gauss-Jordan elimination (GJE) to find out which packet has been newly seen and responds for that packet using the definition and the proposition above. That means the server side can pretend to have received the packet even if it has not been really decoded yet. For example, let us assume the server knows the two linear combinations $c_1=p_1+2p_2+3p_3+4p_4+5p_5$ and $c_2=p_1+p_2+7p_3+8p_4+9p_5$. The server uses GJE to compute $2c_2-c_1=p_1+11p_3+12p_4+13p_5$ and $c_1-c_2=p_2-4p_3-4p_4-4p_5$. According to definition 1 and proposition 1, the linear combinations of $2c_2-c_1$ and $c_1-c_2$ have the form $p_k+q_k$, therefore packets $ p_1$ and $p_2$ are seen, and packets $p_3$, $p_4$ and $p_5$ are unseen. With a large finite field size, every linear combination coming may cause the next unseen packet to be seen. Then, according to theorem $1$ \cite{Jay2011}, if all of the packets in a file have been seen, they can also be decoded.

\subsection{Scenario}
\begin{figure*}[!h]
  \centerline{
  \subfloat[REST HTTP without network coding]{\includegraphics[width=5 in, height=3 cm]{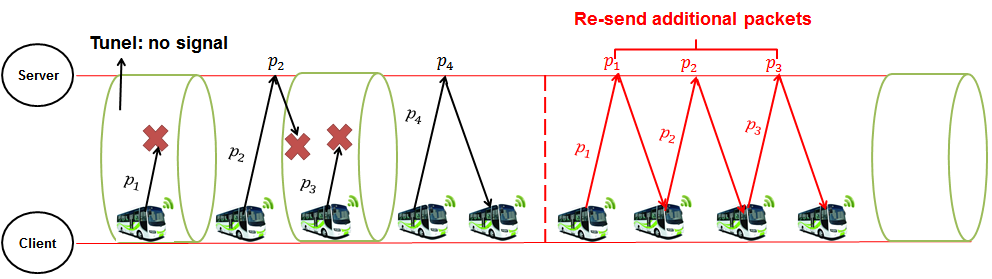}
  \label{scenarioWoNC}}
  }
  \hfil
  \centerline{
  \subfloat[REST HTTP with network coding]{\includegraphics[width=5 in, height=4 cm, scale = 0.8]{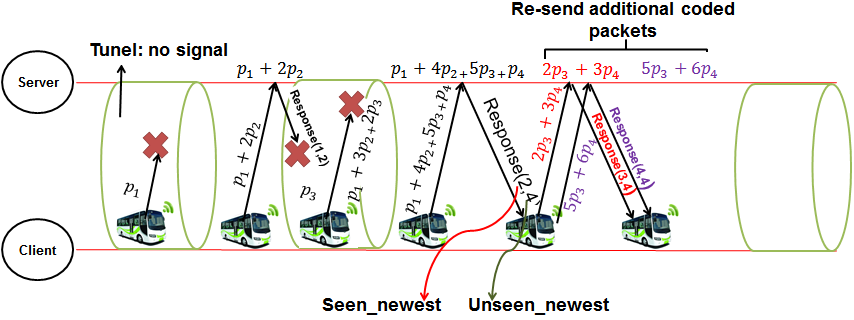}
  \label{scenarioWNC}}
  }
  \caption{Case study of REST HTTP communication with and without network coding}
  \label{scenario}
  \vspace{-0.3cm}
\end{figure*}

As mentioned before, our focus will be on the highly dynamic scenarios. These kind of systems are often met with connectivity and bandwidth issues, causing message losses. We assume REST HTTP based communication and observe the behaviour of particular type of requests. We consider the example, shown in Fig.\ref{scenario}, which takes place between one mobile client (e.g. smart vehicle) and one static server. The client wants to open four connections in order to send four POST request messages, i.e. $p_1, p_2, p_3, and \; p_4$, to the server.

In Fig.\ref{scenarioWoNC} we consider request messages related to unsafe methods exchanged between the client and the server. With REST HTTP, these ones can be safely re-sent several times to receive the responses corresponding to those requests back from the server \cite{Edstrom2012}. However, re-sending them many times while we are not sure what is occurring in unreliable connections, i.e. whether the timeout happened during sending the request to the server or the response to the client, can cause a bandwidth waste in term of the traffic sent.  For example, in the scenario of Fig.\ref{scenarioWoNC}, re-sending message $p_2$ is not necessary because it was already updated at the server side. In order to solve this problem, we propose a RLNC, as shown in Fig.\ref{scenarioWNC}. Before analyzing our scenario, we need to know the two notations: $seen\_newest$  and $unseen\_newest$ contained in response messages from the server module are $IDs$ of the newest seen and unseen message after GJE. Refering to the example of definition $1$ and proposition $1$, after GJE at the server side, we can find out $seen\_newest$ has $ID=2$ which identifies message $p_2$ and $unseen\_newest$ has $ID=5$ belonging message $p_5$.

\begin{figure}[!t]
  \subfloat[]{\includegraphics[width=0.9 in]{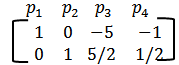}
  \label{matrix1}}
  \hfil \hfil \hfil \hfil \hfil \hfil
  \subfloat[]{\includegraphics[width=0.9 in]{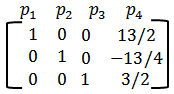}
  \label{matrix2}}
  \hfil \hfil \hfil \hfil \hfil \hfil
  \subfloat[]{\includegraphics[width=0.9 in]{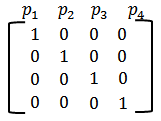}
  \label{matrix3}}
  \caption{Matrices after performing Gauss-Jordan elimination at the server side}
  \label{matrix}
  \vspace{-0.5cm}
\end{figure}

In Fig.\ref{scenarioWNC} we observe that each REST HTTP message is updated at a different time, stored in the NC layer and only removed from the coding buffer when its response is gone back from the server. Request message $p_1$ is lost, therefore at the time of arriving request message $p_2$, a random linear combination of messages $p_1$ and $p_2$ is dispatched to the server, where the coefficients are randomly chosen for the whole message, not each symbol, but its response is lost. Similarly, at the time of arriving $p_3$ and $p_4$, the server has the random linear combinations $p_1+3p_2+2p_3$ and $p_1+4p_2+5p_3+p_4$, respectively, but only the latter is successful on both the client and the server side. At the time of receiving the linear combination $p_1+4p_2+5p_3+p_4$, the server performs GJE on the linear combinations that exist on the server side, and then has the coefficient matrix, as shown in Fig.\ref{matrix1}. With that information in Fig.\ref{matrix1}, the server can respond the response message Response(2,4) containing $seen\_newest=2$ ($ID$ of request message $p_2$) and $unseen\_newest=4$ ($ID$ of request message $p_4$). Note that this response can be sent even when the original request messages have not yet been decoded. Based on Response(2,4), the client can compute by performing $unseen\_newest$ - $seen\_newest$ = $4-2$ = $2$ (2 means the server lacks the two coded messages), and then re-send the two additional random linear combinations (i.e. $2p_3+3p_4$ and $5p_3+6p_4$ ) to compensate losses. The two additional linear combinations do not include request messages $p_1$ and $p_2$ because they have already been removed from the coding buffer after the client received the Response(2,4) (the reason is explained in the part of buffer management at the client side). Fig.\ref{matrix2} shows the coefficient matrix after performing GJE at the time of receiving the additional linear combination $2p_3+3p_4$, where response message Response(3,4) contains $seen\_newest=3$ and $unseen\_newest=4$. Fig.\ref{matrix3} shows the coefficient matrix after performing GJE at the time of receiving the additional linear combination $5p_3+6p_4$, where response message Response(4,4) contains $seen\_newest=unseen\_newest=4$, meaning all original request messages have been decoded. With respect to the message gain, using network coding, we can shorten one resource update cycle compared to the traditional REST HTTP.

However, the problem is still that the current REST HTTP protocol does not allow response to a request message before it has been decoded. Therefore, a modification of REST HTTP is required to respond to every coded request message received by using definition $1$ and proposition $1$ in the paper \cite{Jay2011}. In addition, we use the progressive non-generation-based coding implementation as done for TCP/NC \cite{Jay2011} and dynamic coding \cite{VanVu2014}. On the other hand, as mentioned in our scenario, each request message is updated at a different time, so the newest arrived request is presented by only one linear combination at a time. As a result, to anticipate the number of losses and reasonably adjust the number of additional request messages, we modify the dynamic coding algorithm \cite{VanVu2014} for REST HTTP with network coding.

\subsection{Client NC layer}
\subsubsection{Coding header}
\begin{figure}[htb]
\centering
\includegraphics[width=0.7\columnwidth]{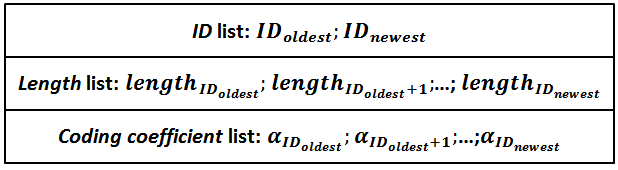}
\caption{Network coding header.}
\label{codingheader}
\vspace{-0.1cm}
\end{figure}

The coding header, shown in Fig. \ref{codingheader}, includes $ID$ list, length of messages list and coding coefficient list involved in the linear combination.

A coded message is generated by forming a linear combination of the messages in the coding buffer, where the coding coefficients are randomly selected for the whole each message, not every symbol. In our implementation data coding is operated over a finite field $\mathbb{F}_{2^8}$. Each message has a specific identifier ($ID$) assigned in order. The header of a coded message will contain information that the server NC layer can use to perform the decoding process and manage its buffer. The meaning of various fields is described as follows.
\begin{itemize}
\item $"ID \; list"$ shows a numbered list of message identifiers involved in a linear combination. $"ID_{oldest}"$ and $"ID_{newest}"$ are the indexes of the oldest and the newest message, buffered into the current coding buffer at the client NC layer. $"ID_{oldest}"$ and $"ID_{newest}"$ are enough in order for the server to know all of the messages in that linear combination. For instance, the client has the linear combination with $"ID_{oldest}"=4$ and $"ID_{newest}"=7$, which means that the linear combination contains $4$ messages $p_4$, $p_5$, $p_6$ and $p_7$, where $p_k$ has the number $ID=k$.
\item $"Length \; list"$ shows the size list of messages and  $length_{i}$ represents the length for the $i^{th}$ message contained in the linear combination. This information is crucial because when implementing the coding process, a problem raises that messages contained in the linear combination have different sizes. In order to address this problem, we may sufficiently append many dummy zero symbols to the shorter messages until all of the messages have the same length. Upon decoding the message at the server NC layer, the dummy zero symbols are pruned using the $"Length \; list"$ header field in the coding header.
\item $"Coding \; coefficient \; list"$ shows the list of coding coefficients and $\alpha_{i}$ denotes the coefficient used for the $i^{th}$ message involved in the linear combination. Note that these ones are randomly chosen for the whole message.
\end{itemize}
\subsubsection{Coding algorithm}
\begin{figure}[htb]
\centering
\includegraphics[width=0.9\columnwidth]{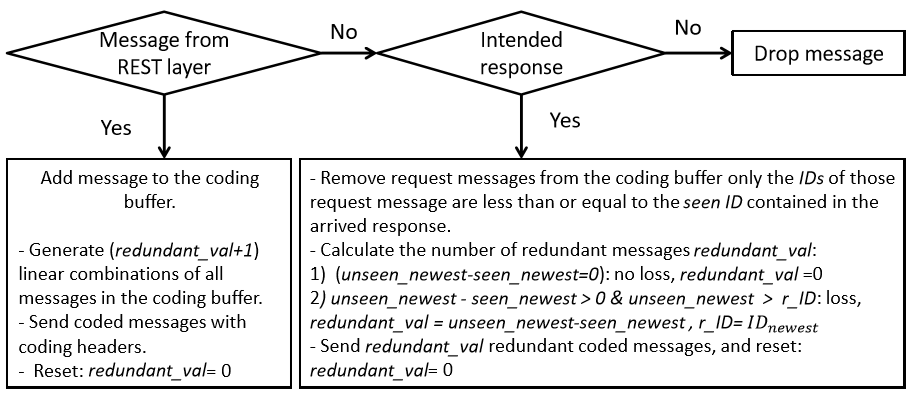}
\caption{Coding algorithm at the client NC layer.}
\label{CodingAlgorithm}
\vspace{-0.1cm}
\end{figure}

This subsection describes the whole coding algorithm on the server NC layer, as shown in Fig.\ref{CodingAlgorithm}. $reduntdant\_val$ is the value denoted for the number of additional messages needed to compensate losses. $r\_ID$ represents the highest $ID$ number of message involved in the additional linear combination. For instance, assume if we re-send the additional linear combination of $3$ messages $p_1$, $p_2$, $p_3$, then $r\_ID$ will be $3$. The operations are detailed as follows.
\begin{itemize}
\item Calculation method for re-sending additional coded messages: The client NC layer accepts messages from the REST layer and stores them into the coding buffer. Then, the client NC layer generates random linear combinations in the coding buffer, some of them including $redudant\_val$ additional linear combinations, where the coding coefficients are randomly chosen for the whole message, and also conveyed in the coding header. Based on $seen\_newest$ and $unseen\_newest$ contained in the response message from the server side, the number of additional coded messages $redundant\_val$ is calculated. If $unseen\_newest$ $-$ $seen\_newest=0$, no loss occurs. Else if $unseen\_newest$ $-$ $seen\_newest$ $>0$ and $unseen\_newest$ $>r\_ID$, then losses occur on the way to the server, therefore we set $redundant\_val$ = $ unseen\_newest$ $-$ $seen\_newest$ and $r\_ID=ID_{newest}$. We reset $redundant\_val=0$ after re-sending the additional messages.
\item Buffer management method: The request messages will be removed from the coding buffer only if the $IDs$ of those request messages are less than or equal to the newest seen $ID$ ($seen\_newest$) contained in the arrived response. If a new request message from the REST layer comes when the buffer is not totally empty, then that one must be dropped and it will be retransmitted later by the REST layer.
\item Subset coding buffer: In case a very small time interval is selected to update information to the server, probably, the client buffers a large number of messages in the buffer. As a result, combining all messages in the coding buffer will make the coding header too large, increasing in that way the coding/decoding complexity. In order to address this problem, we define subset coding buffer that has a fixed size, in order to limit the number of messages in the coding buffer to participate in random linear combinations.
\end{itemize}

\subsection{Server NC layer} \label{DecodingServer}
\begin{figure}[htb]
\centering
\includegraphics[width=0.7\columnwidth]{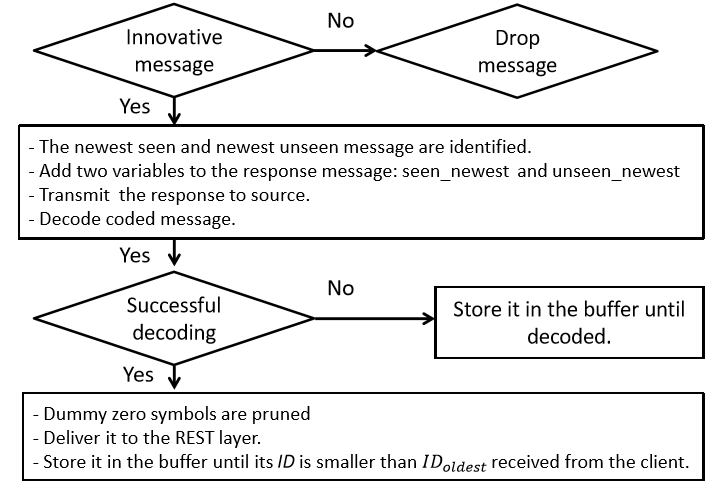}
\caption{Decoding algorithm at the server NC layer.}
\label{DecodingAlgorithm}
\vspace{-0.1cm}
\end{figure}

This subsection describes the whole decoding algorithm on the server NC layer, as shown in Fig.\ref{DecodingAlgorithm}. The operations are detailed as follows.
\begin{itemize}
\item Response method: The server NC layer stores a newly arrived coded message in the decoding buffer, and then reads the coding header and correctly appends the coefficient vector to the decoding matrix. In order to know whether that message is linearly independent, GJE only needs to be performed on the decoding matrix. If the message is not linearly independent, it is deleted. Otherwise, the row transformation operations of GJE on that coded message are also performed. The server NC layer will send a response including the newest seen $ID$ ($seen\_newest$) and newest unseen $ID$ ($unseen\_newest$) identified after GJE, and this job can be performed before the message is decoded and delivered to the REST layer. The seen and unseen $ID$ values are very important for the client NC layer because it uses them to predict and reasonably re-send the number of additional messages.
\item Decoding and delivery method: When a new message is decoded, the dummy zero symbols are pruned using the coding header. After that, that decoded message is delivered to the REST layer.
\item Buffer management method: The arrived coded messages that have not been yet decoded need to be stored in the decoding buffer. The arrived messages without coding or the messages that have been already decoded and delivered are still stored in the buffer until the server NC layer makes sure that they have already been dropped by the client NC layer, then it removes them. This is because they may still be involved in the next linear combinations if their responses are lost on the way to the client side. Using $ID_{oldest}$ belonging to the $"ID\; list"$ header field in the coding header, the server NC layer can remove a decoded message if its $ID$ is smaller than $ID_{oldest}$.
\end{itemize}

\begin{figure*}[ht]
  \centering
  \subfloat[$\alpha=0.3$]{\includegraphics[ width=4.5cm, height=4.4cm]{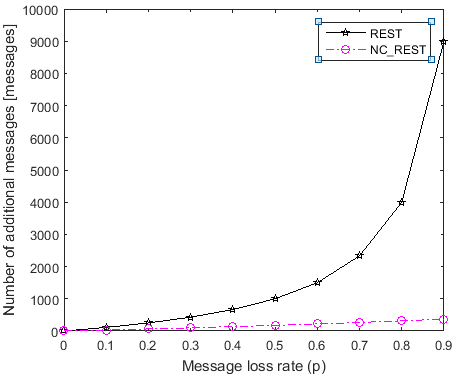}
  \label{1}}
  \subfloat[$\alpha=0.5$]{\includegraphics[ width=4.5cm, height=4.4cm]{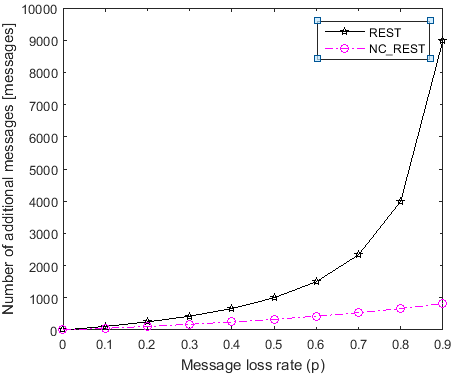}
  \label{2}}
  \subfloat[$\alpha=0.7$]{\includegraphics[ width=4.5cm, height=4.4cm]{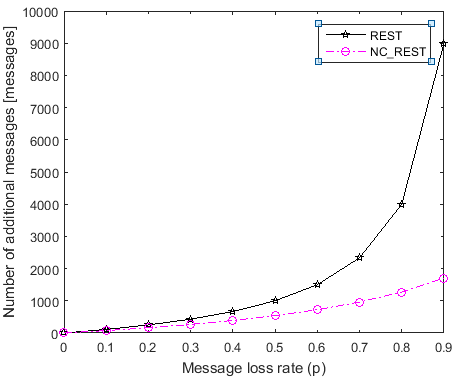}
  \label{3}}
  \subfloat[$\alpha=1$]{\includegraphics[ width=4.5cm, height=4.4cm]{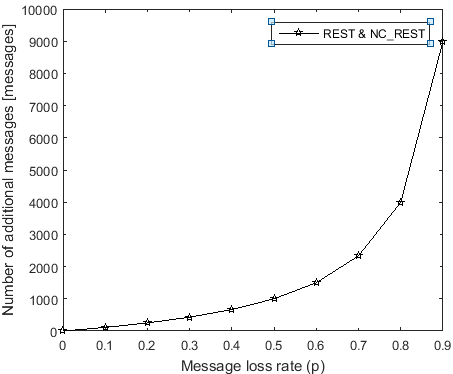}
  \label{4}}
  \caption{Number of additional messages with network coding NC\_REST and without REST}
  \label{additionalmessages}
  \vspace{-0.2cm}
  \end{figure*}

\subsection{Analysis}
We now analyse the impact of REST HTTP with network coding (NC\_REST) on reducing the number of additional messages compared to traditional REST HTTP (REST). Let $p$ be the loss probability including both the loss due to sending request message to the server, and sending response message to the client. $N$ denotes the total number of request messages sent. For REST, re-sending is performed when both lost requests and responses occur, therefore $p$ and $N$, on average, are only fraction $N\cdot(1-p)$ of them successfully negotiated. As a result, to be able to compensate losses for sending $N$ request messages, REST needs to transfer at least $\frac{N}{(1-p)}$ request messages and the number of additional request messages of REST $A_{WoNC}$, which is calculated by:
\begin{equation}
A_{WoNC} = \frac{N}{1-p} - N
\label{A_WoNC}
\end{equation}
where $p<1$. For NC\_REST, re-sending is only considered for lost requests. Let $\alpha$ be the loss rate when sending response message to the client. Hence, to successfully transfer $N$ requests, the number of additional request messages of NC\_REST $A_{WNC}$ is given by:
\begin{equation}
A_{WNC} = \frac{N}{1-(\alpha\cdot p)} - N
\label{A_WNC}
\end{equation}
where $\alpha\cdot p<1$. From Eq.\ref{A_WoNC} and Eq.\ref{A_WNC}, we see that $A_{WNC}\leqslant A_{WoNC}$. We observe that $A_{WNC} = A_{WoNC}$ only when $\alpha=1$, and this is the case where we do not have any benefit from network coding, causing even worse results because it adds additional bytes of overhead for the network coding header in addition to the REST message.

\section{Numerical results}
This section shows numerical results to see the impact of NC\_REST on reducing the number of additional messages and its comparison with REST. For our example, we choose a case with $N=1000$ sent request messages. The message loss probability $p$ including both the request message loss and the response message loss is considered in $[0;0.9]$. Four examples of the request message loss rate $\alpha$ are selected: $\alpha = 0.3$, $\alpha = 0.5$, $\alpha = 0.7$ and $\alpha = 1$.

Fig.\ref{additionalmessages} shows examples of NC\_REST and REST in term of the number of additional messages. The scenarios include different values of $\alpha$ shown in Fig.\ref{1} ($\alpha=0.3$), Fig.\ref{2} ($\alpha=0.5$), Fig.\ref{3} ($\alpha=0.7$) and Fig.\ref{4} ($\alpha=1$). The x-axis and y-axis represent the message loss probability $p$ and the number of additional messages, respectively. The number of additional request messages is calculated by using Eq.(\ref{A_WoNC}) for REST and Eq.(\ref{A_WNC}) for NC\_REST. Observing Fig.\ref{additionalmessages}, the number of additional messages increases when loss probability $p$ increases for both REST and NC\_REST, since the higher loss probability $p$ the more re-sendings occur.

First of all, we consider an example with a small loss rate of $p=0.1$. Compared with NC\_REST, REST increases $259.260\%$, $111.110\%$ and $47.620\%$ for the example shown in Fig.\ref{1}, Fig.\ref{2} and Fig.\ref{3}, respectively. With $p=0.5$, REST needs to re-dispatch $1000$ request messages for all of request message loss rate values $\alpha$, but  NC\_REST only re-sends $176.470$ request messages for $\alpha=0.3$; $333.333$ request messages for $\alpha=0.5$ and $538.461$ request messages for $\alpha=0.7$. In case of a high loss rate of $p=0.9$, REST re-sends $9000$ request messages for all the cases, while NC\_REST only re-sends $369.863$, $818.181$ and $1702.700$ request messages shown in Fig.\ref{1}, Fig.\ref{2} and Fig.\ref{3}, respectively. With those results, NC\_REST always outperforms REST in term of the number of additional messages. Besides that, we see that the lower the loss probability of sending request message to the server $\alpha$ is, the better the benefit of NC\_REST is. The reason for those is because NC\_REST only re-sends for the lost request messages. Therefore, compared to request message loss rate with $\alpha=0.5$, $\alpha= 0.7$ and $\alpha= 1$, a request message loss rate of $\alpha=0.3$ has the best benefit from network coding. For the case of $\alpha=1$ in Fig.\ref{4}, there are no advantages in using NC\_REST, and the number of additional messages for REST and NC\_REST is the same for all loss probability values $p$. Moreover, if we take network coding header into account, NC\_REST will consume an amount of traffic for this, therefore decreasing bandwidth utilization. With the analysed numerical results, we can conclude that NC\_REST outperforms REST in all of cases, except when the request message loss rate is $\alpha=1$.

\section{Conclusion}\label{conclusion}

The network coding method has been used for improving efficiency and bandwidth utilization, as well as the novel paradigm of fog computing. In this paper, taking into consideration highly dynamic scenarios that include the communication between a mobile client and fog processing nodes, where connection is often unreliable, we combine REST HTTP protocol with random linear network coding. We show how our solution can decrease the reconnection time by reducing the additional packet retransmissions. In future works, we will do practical implementation for our algorithm to better understand the impact of network coding on the performance of REST HTTP.

\section*{Acknowledgment}
This work has been partially performed in the framework of mF2C project funded by the European Union's H2020 research and innovation programme under grant agreement 730929.

\bibliographystyle{IEEEtran}
\bibliography{nc-rest}

\begin{thebibliography}{10}
\providecommand{\url}[1]{#1}
\csname url@samestyle\endcsname
\providecommand{\newblock}{\relax}
\providecommand{\bibinfo}[2]{#2}
\providecommand{\BIBentrySTDinterwordspacing}{\spaceskip=0pt\relax}
\providecommand{\BIBentryALTinterwordstretchfactor}{4}
\providecommand{\BIBentryALTinterwordspacing}{\spaceskip=\fontdimen2\font plus
\BIBentryALTinterwordstretchfactor\fontdimen3\font minus
  \fontdimen4\font\relax}
\providecommand{\BIBforeignlanguage}[2]{{%
\expandafter\ifx\csname l@#1\endcsname\relax
\typeout{** WARNING: IEEEtran.bst: No hyphenation pattern has been}%
\typeout{** loaded for the language `#1'. Using the pattern for}%
\typeout{** the default language instead.}%
\else
\language=\csname l@#1\endcsname
\fi
#2}}
\providecommand{\BIBdecl}{\relax}
\BIBdecl

\bibitem{1228459}
T.~{Ho}, R.~{Koetter}, M.~{Medard}, D.~R. {Karger}, and M.~{Effros}, ``The
  benefits of coding over routing in a randomized setting,'' in \emph{IEEE
  International Symposium on Information Theory, 2003. Proceedings.}, 2003, pp.
  442--.

\bibitem{Sharma2017a}
S.~K. Sharma and X.~Wang, ``{Live Data Analytics with Collaborative Edge and
  Cloud Processing in Wireless IoT Networks},'' \emph{IEEE Access}, 2017.

\bibitem{Chiang2016}
M.~Chiang and T.~Zhang, ``{Fog and IoT: An Overview of Research
  Opportunities},'' \emph{IEEE Internet of Things Journal}, 2016.

\bibitem{Vilalta2018}
R.~Vilalta, S.~Via, F.~Mira, R.~Casellas, R.~Munoz, J.~Alonso-Zarate,
  A.~Kousaridas, and M.~Dillinger, ``{Control and Management of a Connected Car
  Using SDN/NFV, Fog Computing and YANG data models},'' \emph{4th IEEE
  Conference on Network Softwarization and Workshops}, 2018.

\bibitem{Gia2018}
T.~{Nguyen Gia}, A.~M. Rahmani, T.~Westerlund, P.~Liljeberg, and H.~Tenhunen,
  ``{Fog Computing Approach for Mobility Support in Internet-of-Things
  Systems},'' \emph{IEEE Access}, 2018.

\bibitem{Li2019}
J.~Li, X.~Shen, L.~Chen, D.~P. Van, J.~Ou, L.~Wosinska, and J.~Chen, ``{Service
  Migration in Fog Computing Enabled Cellular Networks to Support Real-time
  Vehicular Communications},'' \emph{IEEE Access}, 2019.

\bibitem{LeSommer2016}
N.~{Le Sommer}, L.~Touseau, Y.~Maheo, M.~Auzias, and F.~Raimbault, ``{A
  disruption-tolerant RESTful support for the web of things},'' \emph{IEEE 4th
  International Conference on Future Internet of Things and Cloud}, 2016.

\bibitem{Ahlswede2000}
R.~Ahlswede, N.~Cai, S.-y.~R. Li, S.~Member, R.~W. Yeung, and S.~Member,
  ``{Network Information Flow},'' 2000.

\bibitem{Hundeboll2014}
M.~Hundeb{\o}ll, M.~V. Pedersen, D.~E. Lucani, and F.~H. Fitzek, ``{Supporting
  Dynamic Adaptive Streaming over HTTP in wireless meshed networks using random
  linear network coding},'' \emph{International Symposium on Network Coding,
  NetCod 2014 - Conference Proceedings}, 2014.

\bibitem{Peralta2018}
G.~Peralta, R.~Cid-Fuentes, J.~Bilbao, and P.~Crespo, ``{Network Coding-Based
  Next-Generation IoT for Industry 4.0},'' in \emph{Intech open}, 2018.

\bibitem{Marques2017}
B.~Marques, I.~MacHado, A.~Sena, and M.~C. Castro, ``{A Communication Protocol
  for Fog Computing Based on Network Coding Applied to Wireless Sensors},''
  \emph{Proceedings - 29th International Symposium on Computer Architecture and
  High Performance Computing Workshops}, 2017.

\bibitem{Dizdarevic2018}
\BIBentryALTinterwordspacing
J.~Dizdarevic, F.~Carpio, A.~Jukan, and X.~Masip-Bruin, ``{Survey of
  Communication Protocols for Internet-of-Things and Related Challenges of Fog
  and Cloud Computing Integration},'' 2018. [Online]. Available:
  \url{http://arxiv.org/abs/1804.01747}
\BIBentrySTDinterwordspacing

\bibitem{Jay2011}
J.~K. Sundararajan, D.~Shah, M.~M{\'{e}}dard, S.~Jakubczak, M.~Mitzenmacher,
  and J.~Barros, ``{Network Coding Meets TCP: Theory and Implementation},''
  \emph{Proceedings of the IEEE}, vol.~99, no.~3, pp. 490--512, mar 2011.

\bibitem{Edstrom2012}
J.~Edstrom and E.~Tilevich, ``{Reusable and extensible fault tolerance for
  RESTful applications},'' \emph{Proc. of the 11th IEEE Int. Conference on
  Trust, Security and Privacy in Computing and Communications}, 2012.

\bibitem{VanVu2014}
T.~{Van Vu}, N.~Boukhatem, T.~M.~T. Nguyen, and G.~Pujolle, ``{Dynamic coding
  for TCP transmission reliability in multi-hop wireless networks},''
  \emph{Proceeding of IEEE International Symposium on a World of Wireless,
  Mobile and Multimedia Networks 2014, WoWMoM 2014}, pp. 1--6, 2014.

\end{thebibliography}

\end{document}